\documentclass[12pt]{iopart}
\usepackage{iopams}  
\usepackage{graphicx}  
\usepackage{epstopdf} 	
\usepackage{color}
\usepackage{txfonts}

\begin{document}

\title[]{Hybridized exciton-polariton resonances in core-shell nanoparticles}

\author{Martin J Gentile, William L Barnes}
\address{Department of Physics and Astronomy, University of Exeter, Exeter EX4~4QL, UK}
\ead{\mailto{m.j.gentile@exeter.ac.uk}, \mailto{w.l.barnes@exeter.ac.uk}}

\begin{abstract}
The goal of nanophotonics is to control and manipulate light at length scales below the diffraction limit. Typically nanostructured metals are used for this purpose, light being confined by exploiting the surface plasmon-polaritons such structures support. Recently excitonic (molecular) materials have been identified as an alternative candidate material for nanophotonics.  Here we use theoretical modelling to explore how hybridisation of surface exciton-polaritons can be achieved through appropriate nanostructuring. We focus on the extent to which the frequency of the hybridised modes can be shifted with respect to the underlying material resonances.
\end{abstract}

\pacs{7135Gg, 4250Ct, 7135Cc, 7722Ch, 4250Md}

\vspace{2pc}
\noindent{\it Keywords}: nanophotonics, exciton-polariton, J-aggregate, density matrix, core-shell nanoparticle

%
\maketitle
%
%

\vspace{10pt}

\section{Introduction}
Manipulating light beyond the diffraction limit is now an important part of nanoscience and nanotechnology~\cite{Gramotnev_NatPhot_2014_8_13}. Traditionally this has been accomplished by making use of the surface plasmon-polariton modes supported by metallic nanostructures \cite{Barnes_JOptA_2006_8_S87}. These modes allow light to be confined to sub-wavelength volumes, and for the intensity of the associated optical fields to be enhanced \cite{Zayats_PhysRep_2005_408_131}, thereby boosting the strength of light-matter interactions at the nanoscale. There are associated strong field gradients that also offer potential for exploiting chirality at the nanoscale \cite{Hendry_NatNano_2010_5_783,Meinzer_PRB_2013_88_041407(R)}, the production of hot electrons \cite{Brongesma_NatNano_2015_10_25}, and the generation of THz radiation \cite{Welsh_PRL_2007_98_026803,Polyushkin_NL_2011_11_4718}.\\

Recently renewed interest in alternatives to the metals usually used in plasmonics has led to the investigation of a wide range of alternative materials \cite{Naik_AdvMat_2013_25_3264}, for example transparent conducting oxides and alloys. Molecular materials incorporating molecules with excitonic resonances were shown to support surface exciton-polariton (SEP) modes many years ago~\cite{Philpott_MolCrystLiqCryst_1979_50_139}. These modes have many characteristics in common with those of the better-known surface plasmon-polaritons. Surface exciton-polariton modes have been predicted to show field confinement and field enhancement comparable with plasmonic modes \cite{Gentile_NL_2014_14_2339,Gentile_JOpt_2016_18_015001}. The common attribute that metals and excitonic materials possess is that they exhibit a negative permittivity. For metals the negative permittivity is associated with the free (unbound) electrons, the negative response extends over a very wide frequency range, plasmonic systems may thus exhibit great tunability \cite{Jensen_JPhysChemB_2000_104_10549}. For excitonic materials, the bound nature of the exciton means that the negative permittivity is only exhibited over a relatively narrow spectral range~\cite{Philpott_MolCrystLiqCryst_1979_50_139}. Here we wished to explore the extent to which this spectral range could be extended by making use of mode hybridisation~\cite{Prodan_Science_2003_302_419}.\\

The hybridization concept involves the interaction between two modes, it is a concept that has been extensively explored in plasmonics. As an example, a thin metallic film may support a surface plasmon-polariton (SPP) on both of its surfaces. If the media bounding each surface have the same permittivity, then the two SPP modes will be degenerate and if the metal film is thin enough, the fields of the two SPP modes will overlap in the metal. Interaction between the two modes then leads to the formation of two new hybridised states; typically these are referred to as symmetric and asymmetric hybridised modes; the symmetry referring to the charge/field distributions \cite{Sarid_PRL_1981_47_1927,Barnes_OptComm_1986_60_117,Smith_JMO_2008_55_2929}. In the present study we have focussed on the core-shell geometry. The shell of material can be envisaged as a superposition of a particle and a cavity, each with their respective modes: the particle and the cavity mode can hybridise to produce two distinct modes. Hybridisation of the SPP modes associated with more complex plasmonic structures shows the power of this approach for controlling plasmonic modes \cite{Zilio_NL_2015_15_5200}.\\

In the work reported here we explore how the concept of hybridisation may be used to manipulate and modify SEP modes, this has not been done before. It is important to differentiate the work reported here with previous work where excitonic modes have been hybridised with plasmonic modes~\cite{Pockrand_JChemPhys_1979_70_3401}, an area that has been topical in recent years~\cite{Hao_NL_2008_8_3983} especially in the area of strong coupling~\cite{Sugawara_PRL_2006_97_266808,Trugler_PRB_2008_77_115403,Zengin_PRL_2015_114_157401,Torma_RepProgPhys_2015_78_013901,Chikkaraddy_Nature_2016_535_127}.\\

In previous work~\cite{Gentile_JOpt_2016_18_015001}, we used theoretical modelling to demonstrate that polymer nanospheres doped with excitonic dye molecules may support localized surface exciton-polariton (LSEP) modes. However, the single surface topology of the homogeneous nanosphere is somewhat limiting if tuning of the modes is desired. Plasmonic nanoshells are a simple system with a double surface topology that exhibit hybrid modes~\cite{Prodan_Science_2003_302_419,Cacciola_ACSNano_2014_8_11483}: they offer tunability of the modes through adjustment of the nanoshell aspect ratio. Here we make use of a similar nanoshell system. Using theoretical modelling we examine the range of mode tunability one might expect through the use of an excitonic coating on an inert nanosphere, \textit{i.e.} an excitonic nanoshell. We make use of the same dye molecule we considered in our previous work (TDBC, details given below), focussing our attention on the effect of changes to the hybridised modes on: shell thickness, refractive index of the core, and the concentration of dye molecules in the shell. In \Sref{sec:theory} below we outline the two approaches we adopt in our theoretical modelling. In \Sref{sec:res_disc} we present our results before offering conclusions in \Sref{sec:conclusions}.

\section{Theory}\label{sec:theory}

In this work we employ two theoretical approaches. The first is based on a quasistatic analysis. The appeal of this approach is that it offers an analytic model with which we may build some intuitive understanding of the hybridised modes we seek to analyse. Below, we first use this quasistatic model to show how hybridisation occurs. We then switch to a more numerical model based upon a quantum theoretical approach that treats the excitonic units as 4-level quantum systems. Our motivation for switching to this more complex approach is that it provides a better representation of what we should expect from experiment.

\subsection{Quasistatic analysis}\label{sec:secQS}

The quasistatic analysis is based on the frequency-dependent polarizability of the nanostructure. For a small sphere the polarizability in the quasistatic limit is given by~\cite{Barnes_AmJPhys_2016_84_593},

\begin{equation}\label{eq:qs1}
	\alpha = \varepsilon_0\varepsilon_m4\pi r^3\frac{\varepsilon_{\textrm{s}}-\varepsilon_m}{\varepsilon_{\textrm{s}}+2\varepsilon_m},
\end{equation}

\noindent where $r$ is the radius of the particle, $\varepsilon_0$ is the permittivity of free space, $\varepsilon_{\textrm{s}}$ is the relative permittivity of the material from which the sphere is made, and $\varepsilon_m$ is the permittivity of the material surrounding the nanoparticle. The quasistatic polarizability can be written in a similar form for the core-shell particle~\cite{Fofang_NL_2008_8_3481},

\begin{equation}\label{eq:qs2}
	\alpha = \varepsilon_0\varepsilon_m4\pi r_2^3\frac{\varepsilon_{\textrm{eff}}-\varepsilon_m}{\varepsilon_{\textrm{eff}}+2\varepsilon_m},
\end{equation}

\noindent where now $r_2$ is the radius of the combined core-shell particle and $\varepsilon_{\textrm{eff}}$ is the effective permittivity of the core-shell system, discussed below.\\

\begin{figure}[h]
	\centering
	\includegraphics[width=0.3\columnwidth]{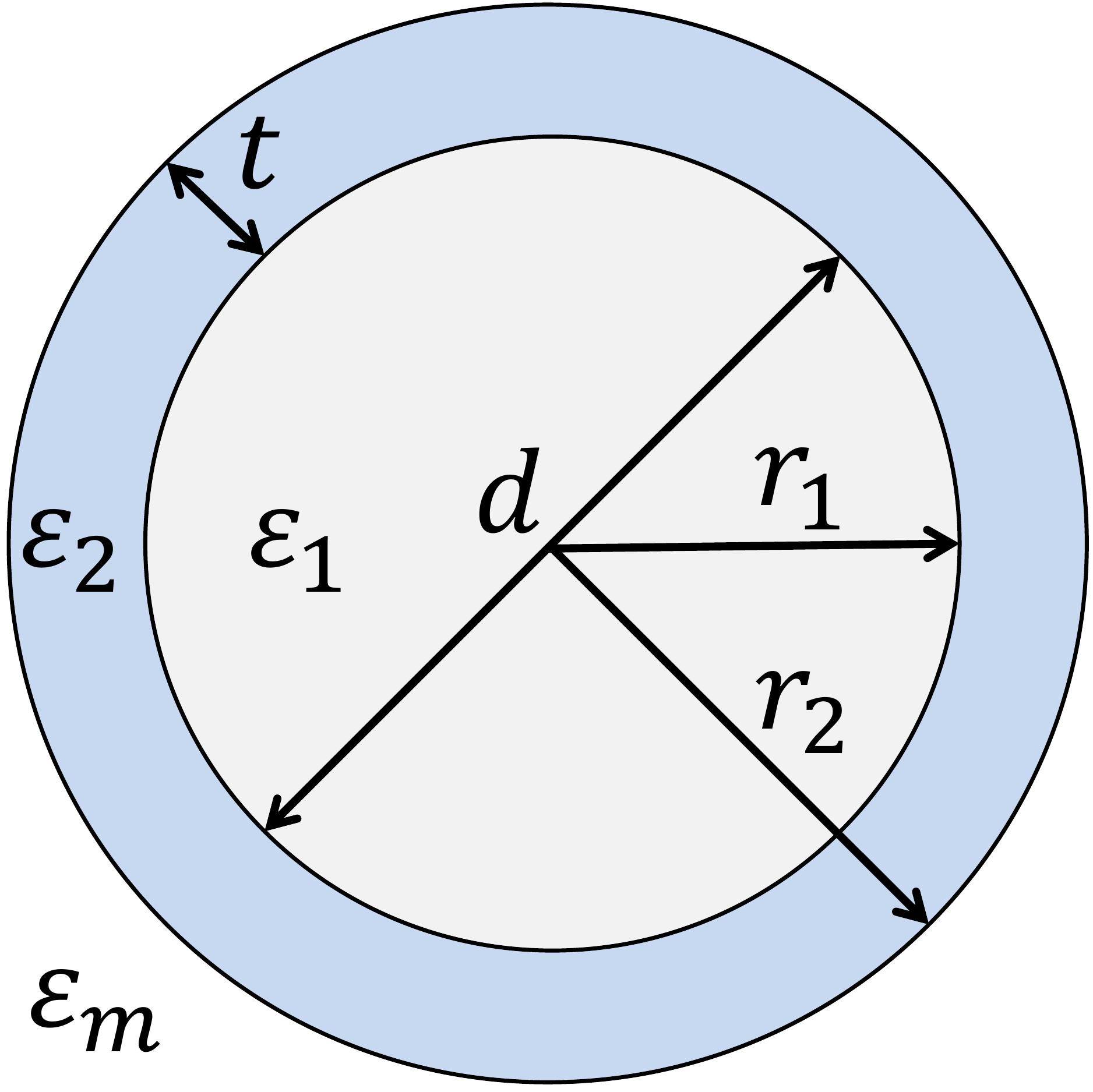}
	\caption{Schematic of the coated nanosphere. The core, of radius $r_1$ and diameter $d$ is an inert dielectric of relative permittivity $\varepsilon_1$. The shell, of thickness $t$ is the excitonic region and has a frequency-dependent relative permittivity of $\varepsilon_2$. The outer radius is $r_2$ and the relative permittivity of the surrounding medium is $\varepsilon_m$.}
	\label{fig:CoatedNanosphere}
\end{figure}

\Fref{fig:CoatedNanosphere} shows a schematic of the core-shell structure. The coated nanosphere has a core of diameter $d$, an inner radius $r_1$, an outer radius $r_2$, and a coating (nanoshell) thickness $t$. The relative permittivities of the core and the shell are denoted $\varepsilon_1$ and $\varepsilon_2$ respectively. The relative permittivity of the surrounding medium is again written as $\varepsilon_m$.
  
Resonant modes of the core-shell system are associated with minima in the denominator of \Eref{eq:qs2}. To find the frequencies at which the modes (minima) occur we make use of the effective permittivity of a core-shell system in the quasistatic limit which can be written as~\cite{Fofang_NL_2008_8_3481},

\begin{equation}
	\varepsilon_{\textrm{eff}} = \varepsilon_2\frac{2\varepsilon_2+\varepsilon_1-2x^3(\varepsilon_2-\varepsilon_1)}{2\varepsilon_2+\varepsilon_1+x^3(\varepsilon_2-\varepsilon_1)},
	\label{eq:epsilon_eff}
\end{equation}

\noindent where the aspect ratio of the nanoshell, $x$, is defined as $x\!=\! r_1/r_2$. Substitution of \Eref{eq:epsilon_eff} into the denominator of \Eref{eq:qs2} yields the following expression,

\begin{equation}
	2(1-x^3)\varepsilon_2^2+[(1+2x^3)\varepsilon_1+2\varepsilon_m(2+x^3)]\varepsilon_2+2\varepsilon_m\varepsilon_1(1+x^3)=0.	
	\label{eq:general1}
\end{equation}

\noindent The frequencies for which resonance occurs for given values of $x$, and for a given combination of core-shell materials are found from the solutions to this equation. Since \Eref{eq:general1} is quadratic in $\varepsilon_2$, solutions for $\varepsilon_2$ can be written as,

\begin{equation}
	\varepsilon_2=\frac{-[(1+2x^3)\varepsilon_1+2\varepsilon_m(2+x^3)]\pm\sqrt{A}}{4(1-x^3)},
\end{equation}
  
\noindent where,

\begin{equation}
	A=[(1+2x^3)\varepsilon_1+2\varepsilon_m(2+x^3)]^2-16\varepsilon_m\varepsilon_1(1-x^6).
	\label{eq:mastereq}
\end{equation}

\noindent In other words, for a given $\varepsilon_1$, $\varepsilon_m$ and $x$, resonances occur when the frequency is such that $\varepsilon_2$ enables \Eref{eq:general1} to be satisfied. For inert cores, $\varepsilon_1$ can be assumed to be independent of frequency. This assumption is a good approximation for many materials and permits analytical solutions for the resonant frequencies to be found - this assumption is made for the core in this work. For the shell, it is assumed that $\varepsilon_2$ can be modelled as a Lorentz oscillator dielectric, \textit{i.e.,}

\begin{equation}
	\varepsilon_2 = \varepsilon_\infty + \frac{f\omega_0^2}{\omega_0^2-\omega^2-i\gamma\omega},
	\label{eq:lorentz}
\end{equation}

\noindent where $f$ is the reduced oscillator strength, $\omega_0$ is the central (angular) frequency of the resonance, $\omega$ is the angular frequency, and $\gamma$ is the damping rate of the resonance. Substitution of \Eref{eq:lorentz} into \Eref{eq:mastereq} with re-arrangement for $\omega$ gives two solutions for the resonant frequencies ($\omega$) in the limit that $\gamma\ll\omega$; these can be written as~\cite{Gulen_JPCB_2013_117_11220},

\begin{equation}
	\omega^2_{\pm} = \omega_0^2+\frac{4\omega_0^2f(1-x^3)}{4(1-x^3)\varepsilon_\infty+[(1+2x^3)\varepsilon_1+2(2+x^3)\varepsilon_m]\mp\sqrt{A}}.
	\label{eq:omega_solns_coreshell}
\end{equation}

The form of \Eref{eq:omega_solns_coreshell} makes the result of hybridisation evident, and resonance occurs at two distinct positive frequencies, $\omega_{\pm}$ (the two negative frequencies implied by \Eref{eq:omega_solns_coreshell} correspond to those obtainable through time-reversal symmetry, and are not relevant to the present situation).\\

\begin{figure}[h]
	\centering
	\includegraphics[width=0.5\columnwidth]{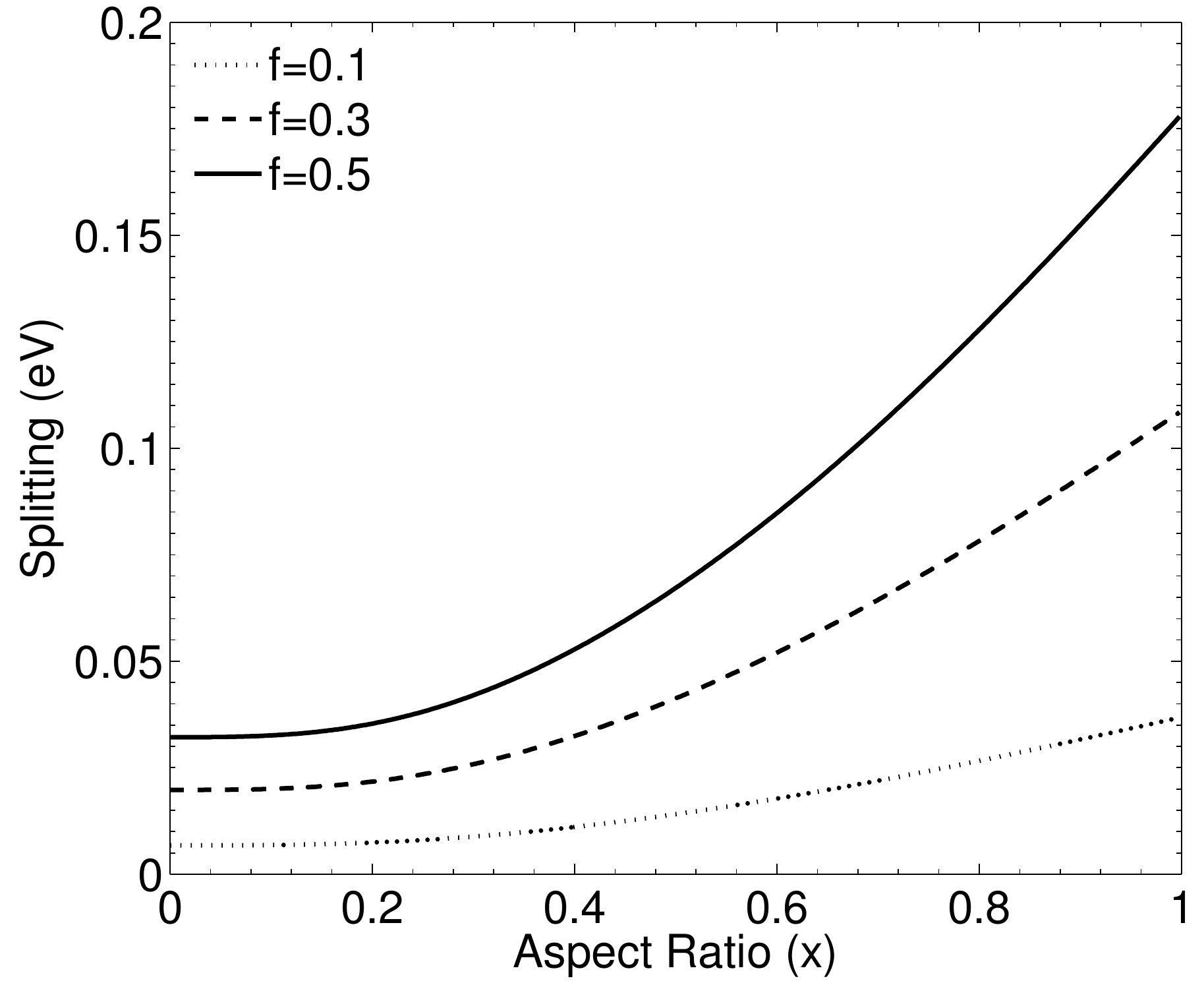}
	\caption{Splitting ($\omega_+-\omega_-$) as a function of aspect ratio $x$ for three oscillator strengths $f$, using \Eref{eq:omega_solns_coreshell} with $\omega_0=2.11~eV$, $\varepsilon_\infty=2.1316$ and $\varepsilon_1=2.25$.}
	\label{fig:SplittingTDBC_xf}
\end{figure}

\noindent For an increase in dye concentration, the oscillator strength ($f$) in \Eref{eq:omega_solns_coreshell} is increased, thereby increasing the splitting of the resonances by a factor of $\sqrt{1+kf}$, where $k$ is a constant, as illustrated in \Fref{fig:SplittingTDBC_xf}. Also shown in the figure is the dependency upon nanoshell aspect ratio: for thin shells ($x\rightarrow 1$), the splitting increases up to a finite limit, a result which can be proven by application of l'H\^{o}pital's rule. Therefore, it is expected that the largest splittings will be occur for thin nanoshells with high dye concentrations.\\

If we relax the assumption that $\gamma\ll\omega$, i.e. we seek solutions when $\gamma\nll\omega$, then we obtain the following (more precise) equation for the resonant frequencies,

\begin{equation}
	c_4\omega^4+c_2\omega^2+c_0=0,	
	\label{eq:quadratic_solns}
\end{equation}

\noindent where,\\

$c_4 = \varepsilon_1\varepsilon_\infty(2x^3+1)+2(\varepsilon_1\varepsilon_m+\varepsilon_\infty^2)(1-x^3)+2\varepsilon_\infty\varepsilon_m(x^3+2)$,\\

	$c_2 = -\varepsilon_1(\varepsilon_\infty(\gamma^2+2\omega_0^2)+f\omega_0^2)(2x^3+1)-2\varepsilon_\infty\varepsilon_m(\gamma^2+2\omega_0^2(1-\varepsilon_mf))(x^3+2)+2(\varepsilon_1\varepsilon_m(\gamma^2+2\omega_0^2)+\varepsilon_\infty^2\gamma^2+4\varepsilon_\infty\omega_0^2(\varepsilon_\infty+f))(x^3-1)$,\\

	$c_0 = \varepsilon_1\omega_0^4(\varepsilon_\infty+f)(2x^3+1)+2\omega_0^4(\varepsilon_1\varepsilon_m+\varepsilon_\infty^2+2\varepsilon_\infty f+f^2)(1-x^3)+2\varepsilon_m\omega_0^4(\varepsilon_\infty+f)(x^3+2)$.\\

\noindent In short, by analysing nanoshell systems with a quasistatic treatment, one can predict hybridization of particle modes with relative ease, and anticipate their change in behaviour with respect to the environment and the intrinsic material properties of the core and nanoshell. Where quasistatic solutions are employed in the remainder of this paper \Eref{eq:quadratic_solns} is used.

\subsection{Quantum model}\label{sec:secQM}

The Lorentz model for the permittivity, whilst useful for building intuition, is not ideal for making quantitative scattering and absorption calculations: the approximate semi-classical nature of such a model is rather restrictive. In a previous work we made use of a quantum approach to better simulate the response of the excitonic material~\cite{Gentile_JOpt_2016_18_015001}. In that work we made use of a two-level quantum model for the permittivity, using the Liouville-von Neumann equation~\cite{Blum_1996} with constant terms in the Lindblad superoperator~\cite{Schirmer_PRA_2004_70_022107,Breuer_2002} to account for damping due to population decay and dephasing effects in the material. In the present work, we adopt a slightly more refined model based on a four-level linear quantum model fitted at optical frequencies to experimentally-determined values of the relative permittivity for four different dye concentrations~\cite{Gentile_NL_2014_14_2339}.\\

We chose to model TDBC-doped polyvinyl alcohol (PVA) as a nanoshell material. This molecule (TDBC) is \textit{5,6-dichloro-2-[[5,6-dichloro-1-ethyl-3-(4-sulphobutyl)-benzimidazol-2-ylidene]-propenyl]-1-ethyl-3-(4-sulphobutyl)-benzimidazolium hydroxide}, \textit{sodium salt, inner salt}; and was chosen because of its tendency to form J-aggregates. Aggregation leads to a narrowing of the optical resonance associated with the excitonic transition in this molecule \cite{van_Burgel_JChemPhys_1995_102_20,Bradley_AdvMat_2005_17_1881}, to the extent that, at sufficiently high concentrations, materials doped with such molecules exhibit a negative permittivity. It is this negative permittivity that enables these materials to support localized resonances.\\

Here we use a similar quantum approach to that used in our previous work, but have extended it by refining our model from a two-level model to a four-level linear quantum model. Doing so gives us a better match between the model and the experimental data for the permittivity of the TDBC:PVA across optical frequencies.\\

The set of energy eigenstates ($|m\rangle$) in the single-exciton regime for an aggregate of $n$ monomers is~\cite{Malyshev_PRB_1995_51_14587,Miura_2012},

\begin{equation}
	|m\rangle=\sqrt{\frac{2}{n+1}}\sum^n_{j=1}\sin\left(\frac{jm\pi}{n+1}\right)|1_j\rangle,
	\label{eq:excited_state_gen}
\end{equation}

\noindent where $|1_j\rangle$ represents the exciton at site $j$. The transition dipole moment of each state $m$ of the aggregate can be derived from \Eref{eq:excited_state_gen} and is given by~\cite{van_Burgel_JChemPhys_1995_102_20,Hochstrasser_JCP_1972_56_5945},

\begin{equation}
	\bi{d}_{01}(m) = \boldsymbol{\mu}_{01}\sqrt{\frac{1-(-1)^m}{n+1}}\cot\left (\frac{\pi m}{2(n+1)}\right ),
	\label{eq:transition_dipole_moment}
\end{equation}

\noindent where $\boldsymbol{\mu}_{01}$ is the transition dipole moment of each monomer (assumed identical). Given that transitions between states above the ground state are forbidden, \Eref{eq:transition_dipole_moment} implies that the first few transitions between the ground state and odd eigenstates are chiefly responsible for the optical response of the material. The transitions between the ground state and even states are `dark' (where a `dark' state has a zero transition dipole moment~\cite{Knoester_2002}). The eigenvalues of the eigenstates in \Eref{eq:excited_state_gen} are written in the following way,

\begin{equation}
	\hbar\omega_1 = \hbar\omega_1^{(1)}-2J\cos\left (\frac{\pi}{n+1}\right ),
\end{equation}

\noindent where $\hbar\omega_1^{(1)}$ is the exciton transition energy of the monomer, and $J$ is the coupling strength between the monomers within the aggregate (the inter-aggregate interactions are assumed to be small, and are neglected). In this paper, four eigenstates are considered: the ground state, and the first three `bright' states (where $\bi{d}_{01}\neq 0$). The rationale for restricting ourselves to these four states is that we are interested in the response of nanoparticles at visible frequencies; the corresponding wavelength of the transition between the highest state considered ($|5\rangle$) and the ground state ($|0\rangle$) is $463.5~nm$, whereas the wavelength of the next such transition ($|0\rangle-|7\rangle$) is $393.0~nm$, outside the visible range. The fundamental parameters used in the equations above are $J=666.5~meV$, $\hbar\omega_{01}^{(1)}=3.417~eV$, and $\mu_0=20~D$. The dephasing parameters~\cite{Gentile_JOpt_2016_18_015001} used in the model (following the four-level quantum constraints~\cite{Schirmer_Solomon_PRA_70_022107_2004}) are $\Gamma_{01}^{(d)}=18.8~meV$, $\Gamma_{03}^{(d)}=\Gamma_{05}^{(d)}=75.1~meV$. By holding the host environment to be constant, these dephasing parameters are also constant. This theoretical treatment leaves the number of monomers \textit{per} aggregate ($n$) as the one remaining free parameter. By using a best-fit procedure to fit the theoretical values for $\varepsilon$ to the experimentally-determined values, $n$, for each of the four dye concentrations was determined. By this process, the following plausible dependency of $n$ upon concentration was found,

\begin{equation}
	n=\left\lfloor aCe^{-bC}+n_0\right\rceil, 
\end{equation}

\noindent where $C$ is the fraction of dye present by weight, $n_0=15$, and the values of the dimensionless constants $a$ and $b$ have been determined as $a=10.6\times 10^5$ and $b=704$. For very low concentrations, $n$ is small. For a single monomer present, $n$ is unity. As the concentration increases, $n$ peaks, before tailing off towards $n_0$ for high concentrations.\\

\begin{figure}[h]
	\centering
	\includegraphics[width=0.5\columnwidth]{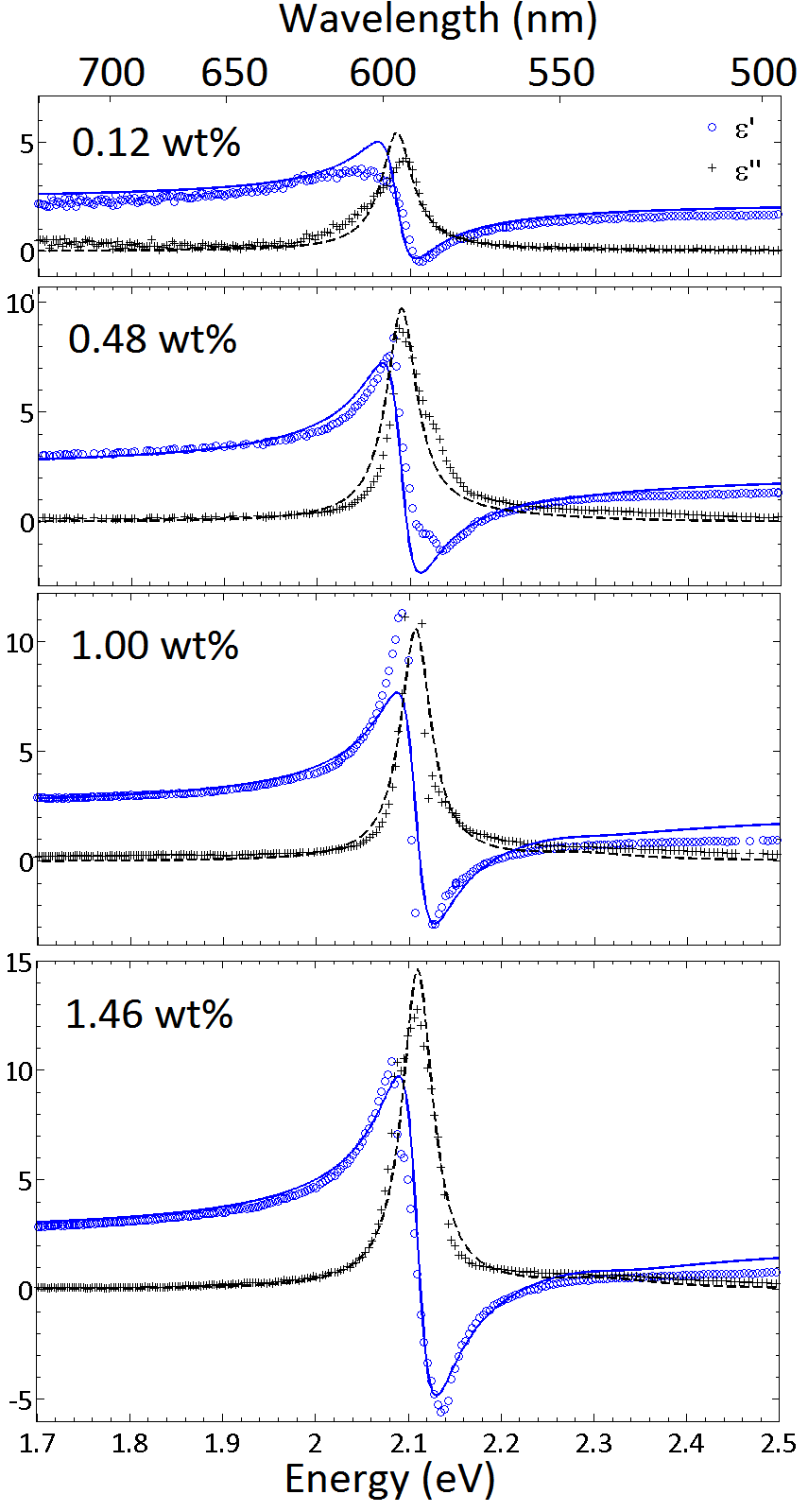}
	\caption{The experimental (circles and crosses) and quantum-mechanically modelled (lines) complex permittivity of TDBC:PVA films for concentrations of (top to bottom) $0.12~wt\%$, $0.48~wt\%$, $1.00~wt\%$ and $1.46~wt\%$. The values of $n$ used for these concentration are 70, 31, 16 and 15 respectively. The other parameters used are outlined in the main text. Blue indicates the real part of the permittivity, black the imaginary part.
	\label{fig:PermittivityTDBC}}
\end{figure}

\noindent The parameters $a$ and $b$ were determined using a least-squares fit procedure to the values of $n$ determined by the best fit to the experimental data for the permittivity. The results for the permittivity (and values of $n$) of TDBC:PVA for the four different dye concentrations are displayed in \Fref{fig:PermittivityTDBC}.\\

\noindent The data displayed in \Fref{fig:PermittivityTDBC}  show several interesting features. First, the real part of the permittivity goes negative on the high energy side of the resonance. Second, the extent of this negativity and the spectral width over which it occurs increases with increasing dye concentration. For the lowest concentration (\Fref{fig:PermittivityTDBC}, top panel) the permittivity is not sufficiently negative to reach $\varepsilon_{\rm{S}}<-2$, the least negative value of the permittivity for which a localised resonance may occur, the Fr\"{o}hlich condition~\cite{Frohlich_1963}. Indeed, achieving a real part of the permittivity below this value only occurs in the data shown for a dye concentration of $1~wt\%$ and above. For the highest concentration for which data are shown ($1.46~wt\%$), the real part of the permittivity is below $-2$ in the energy interval $2.12~eV<E<2.17~eV$. The resonance of a solid nanosphere of this material may only occur in this range, although we note that other shapes may change this situation~\cite{Barnes_AmJPhys_2016_84_593}.\\

\begin{figure}[h]
	\centering
	\includegraphics[width=0.5\columnwidth]{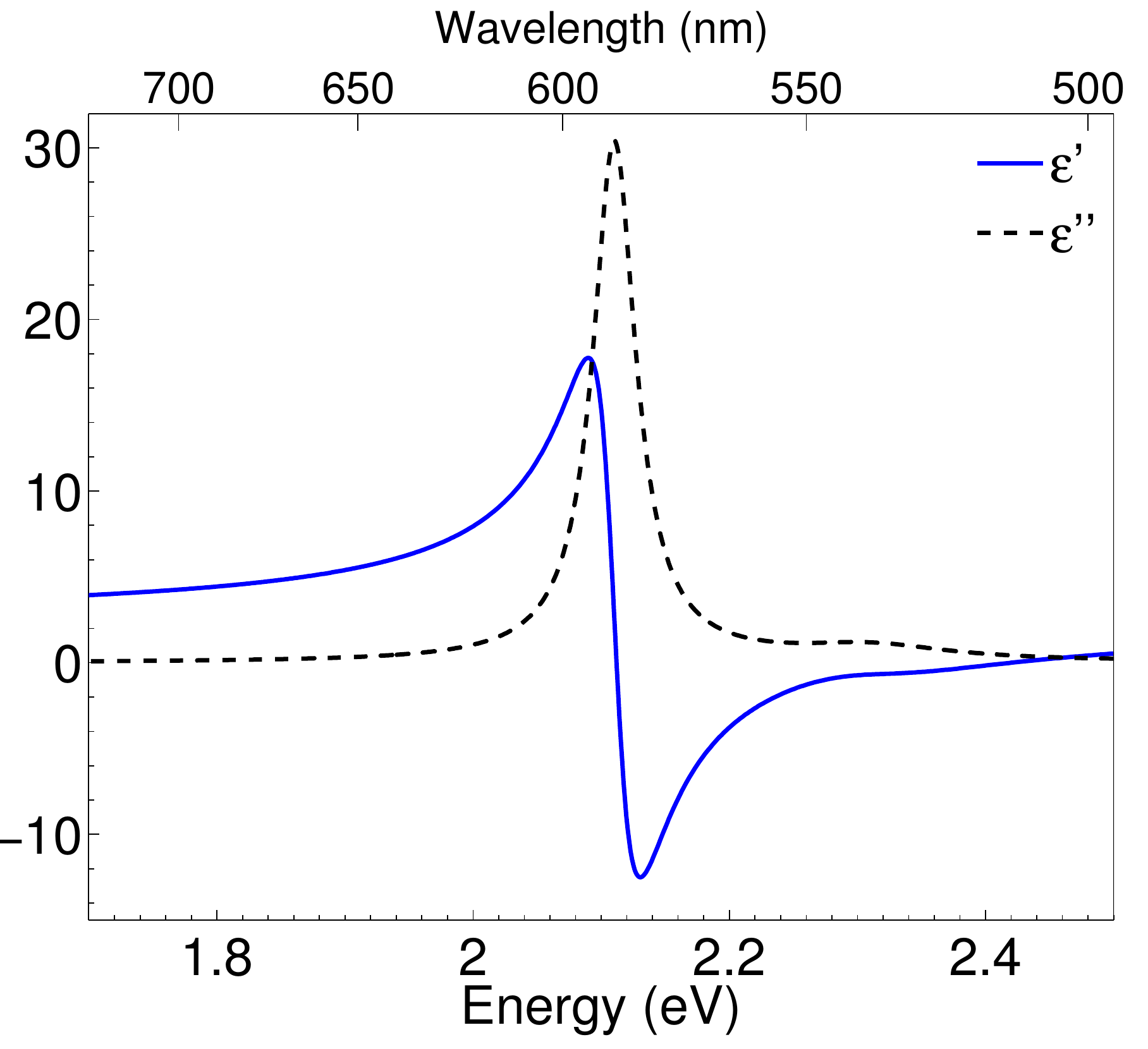}
	\caption{The calculated real (solid line) and imaginary (dashed line) parts of the relative permittivity for $3~wt\%$ TDBC:PVA. Calculations were carried out using the four-state quantum model.}
	\label{fig:TDBC3pc}
\end{figure}

This improved quantum model for permittivity accompanied by the fit for the number of monomers \textit{per} aggregate now gives us the freedom to compute the permittivity of TDBC:PVA for arbitrary molecular concentrations. Doing so for a dye concentration of $3~wt\%$ results in the values shown in \Fref{fig:TDBC3pc}. Here, the range of photon energies for which $\varepsilon'\!<\!-2$ is $2.11~eV<E<2.24~eV$: an expansion of $0.8~eV$ on that obtainable with the $1.46~wt\%$ sample.

We calculated the absorption efficiency of nanospheres with inert cores and TDBC:PVA shells using a core-shell Mie code adapted from the code provided by Bohren and Huffman~\cite{BandH}. Absorption efficiency is defined as ($Q_{abs}=\sigma_{abs}/\pi r_2^2$). We chose to calculate $Q_{abs}$ since this quantity enables relatively easy deduction of anomalously strong absorption features. In particular, we are interested in the case where $Q_{abs}>1$, i.e. where the nanoparticle absorption is more than its geometrical cross section. This property is indicative of nanoparticle plasmon modes in the field of plasmonics~\cite{Hohenau_2007}, and can be used in a similar way for LSEP modes.\\

\section{Results and Discussion}\label{sec:res_disc}

\begin{figure}[h]
	\centering
	\includegraphics[width=0.5\columnwidth]{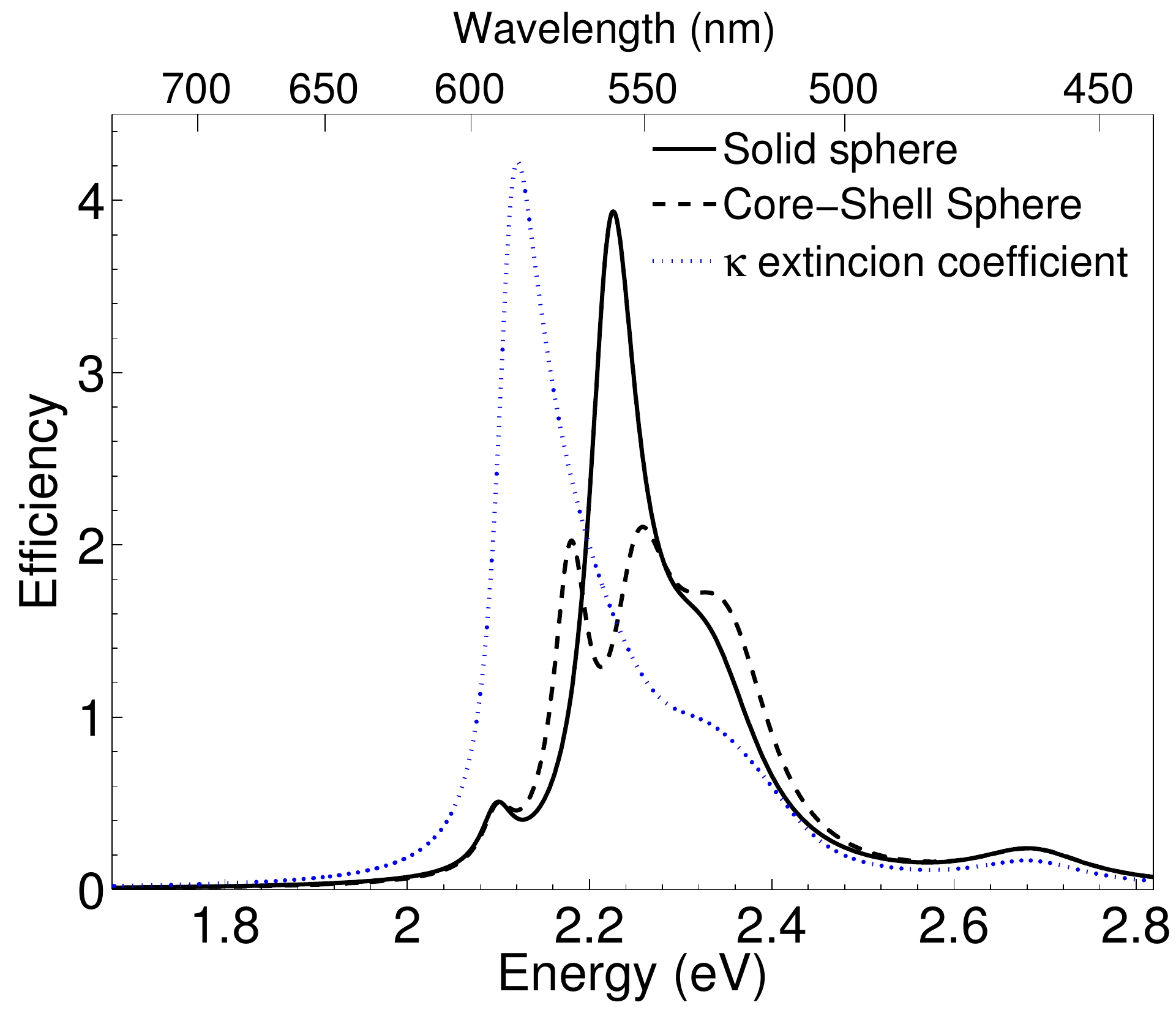}
	\caption{$Q_{abs}$ (black solid line) for a $40~nm$ radius $3.00~wt\%$ TDBC:PVA nanosphere. The black dashed line shows $Q_{abs}$ for a $20~nm$ radius $TiO_2$ core with a $20~nm$ thick shell of $3.00~wt\%$ TDBC:PVA. In both cases the surrounding material is vacuum. The blue dotted line depicts the imaginary part of the refractive index for the shell material. Calculations were carried out using the four-state quantum model.}
	\label{fig:NewFig3}
\end{figure}

In \Fref{fig:NewFig3}, we show the calculated absorption efficiency for a solid $3~wt\%$ TDBC:PVA nanosphere of radius $40~nm$. Also shown is the absorption efficiency for a core-shell nanoparticle where the $20~nm$ thick shell is 3\% TDBC:PVA, and has a $20~nm$ radius core taken to be made of $TiO_2$. The frequency-dependent permittivity of $TiO_2$ used here was taken from the literature~\cite{Conrady_1960}. For comparison we also show the extinction coefficient for the $3~wt\%$ TDBC:PVA. We chose the core material of the core-shell nanoparticle to be $TiO_2$ for several reasons: first, it has low loss, i.e. it has a loss tangent ($\varepsilon''/\varepsilon'$) of approximately zero, which enables clear illustration of the two resonances; second, the two modes are of roughly equal strength (this is discussed in more detail in the next section); and third, for the suitability of $TiO_2$ for nanosphere fabrication~\cite{Li_AdvFuncMat_2011_21_1717}, thereby providing a practical nanoparticle system for experimental investigation. 

For the solid nanosphere the peak absorption efficiency occurs at a higher energy than the peak extinction coefficient, as expected from a localised surface exciton-polariton mode, as noted in previous work~\cite{Gentile_JOpt_2016_18_015001}. The response of the core-shell nanosphere is very different. Now we see that there are two peaks in $Q_{abs}$. The two modes indicated by these peaks are the hybridised modes associated with the core-shell structure. Confirmation of this interpretation comes from examining the dependence of these modes on the thickness of the shell. We anticipate that as the shell thickness is reduced the spectral splitting of the modes will increase, owing to the increased interaction between the two `bare' modes, \textit{i.e.} the SEPs associated with the outer and inner surfaces of the shell.

\subsection{Shell thickness}

\begin{figure}[h]
	\centering
	\includegraphics[width=0.5\columnwidth]{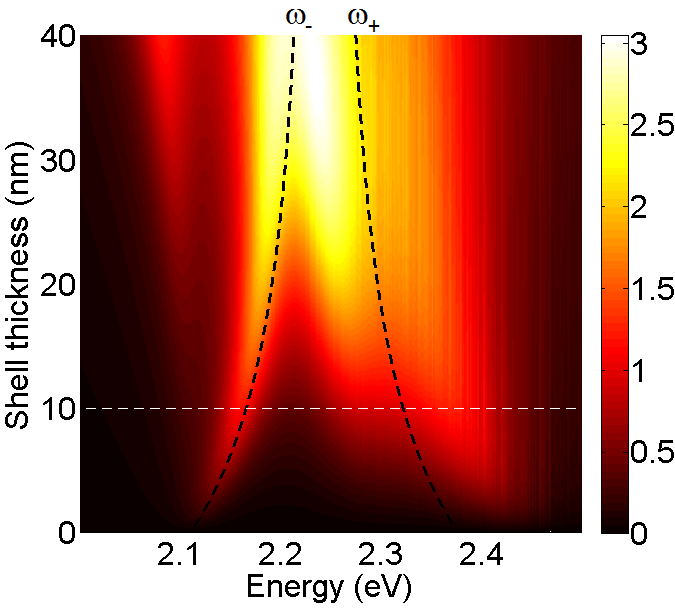}
	\caption{$Q_{abs}$ for a $40~nm$ diameter $TiO_2$ core with a $3.00~wt\%$ TDBC:PVA coating. Calculations were carried out using the four-state quantum model. Black dashed lines show resonant modes obtained from the quasistatic Lorentz oscillator model using \Eref{eq:quadratic_solns}. The white dashed line indicates the shell thickness used in \Fref{fig:d40nm_two_ts_TDBC_3pc_n1_8_yz}c~\&~d.}
	\label{fig:TDBC_d40nm_n1_8_3p_tvary}
\end{figure}

In \Fref{fig:TDBC_d40nm_n1_8_3p_tvary} we show calculated $Q_{abs}$ spectra as a function of shell thickness in the form of a colour plot. Data are presented for a $d=40~nm$ core of $TiO_2$ . Again we chose $3.00~wt\%$ TDBC:PVA as the shell material. 

\noindent Also shown in \Fref{fig:TDBC_d40nm_n1_8_3p_tvary} are the spectral positions of the hybridized modes calculated using the analytical theory from \Sref{sec:secQS}, they are shown as black dashed lines. For the analytical model (\Eref{eq:quadratic_solns}) we used a fixed value for the permittivity of the core, taking that value to be the mean over the optical range, $\varepsilon_1=6.43$.\\

From the data shown in~\Fref{fig:TDBC_d40nm_n1_8_3p_tvary} we can see that as the thickness of the shell in reduced the splitting of the two hybridized modes increases, and the modes become weaker - as expected due the decreasing volume of excitonic material involved. The strength of the splitting reaches approximately $0.2~eV$, roughly 10\% of the central frequency. Considered from the perspective of the width of the imaginary part of the relative permittivity of the $3.00~wt\%$ TDBC:PVA, approximately $0.05~eV$ (see \Fref{fig:PermittivityTDBC}, lower panel), the extent of the splitting is very significant. 
Whilst the extent of the splitting cannot match that possible with plasmonic systems, it does indicate that a very considerable degree of spectral control is possible for hybridised modes, despite the fixed nature of the excitonic resonances upon which they are based. 

The positions of the resonances in \Fref{fig:TDBC_d40nm_n1_8_3p_tvary} predicted by the analytical model provide a reasonable match to the quantum mechanical calculations. The regions within which $Q_{abs}>1$ indicate where the nanoparticle absorbs more light than is geometrically incident upon it, i.e. where the nanoparticle acts like a magnet for light. \Fref{fig:TDBC_d40nm_n1_8_3p_tvary} shows that maximum splitting is achieved for very thin shells. This increase in splitting comes at the cost of a loss in the peak values of $Q_{abs}$, due to the reduction in the volume of the excitonic material.

\begin{figure}[h]
	\centering
	\includegraphics[width=0.75\columnwidth]{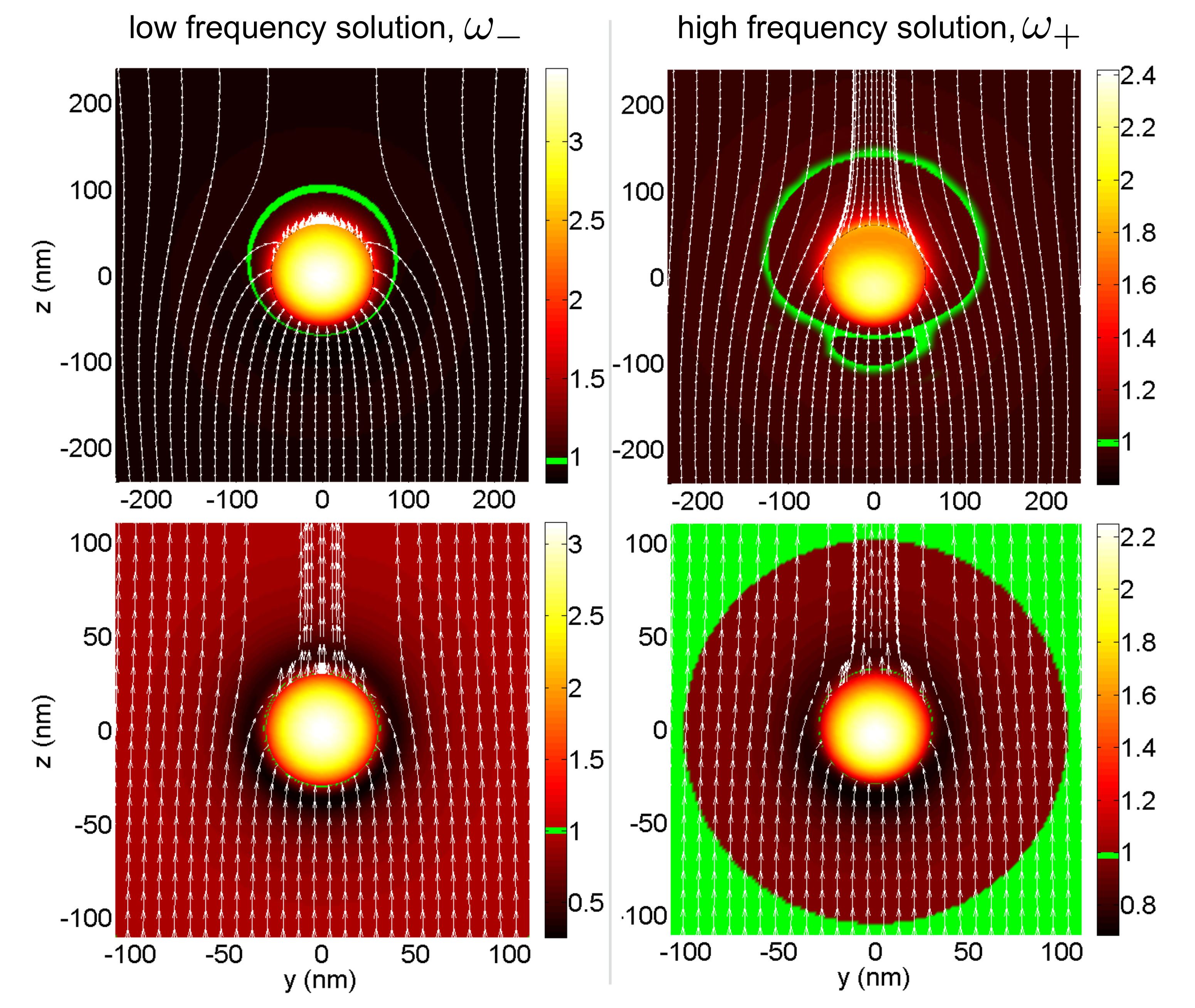}
	\caption{Plots on peak resonance in the y-z plane for nanospheres with a $TiO_2$ core of $d=40~nm$ and coatings of $3.00~wt\%$ TDBC:PVA with shell thicknesses (upper row) $40~nm$ and (lower row) $10~nm$. Left-hand column plots are for the higher frequency $(\omega_-)$ mode, the right-hand column plots are for the lower frequency $(\omega_+)$ modes. Colour plot: time-averaged magnitude of the electric field, where the green line bounds the region within which there is field enhancement. Streamlines: power flow, from the total Poynting vector around the nanospheres. The resonant energies used in plots a, b, c and d are $2.21~eV$ ($560.6~nm$), $2.28~eV$ ($545.4~nm$), $2.17~eV$ ($572.8~nm$) and $2.32~eV$ ($534.3~nm$) respectively. Calculations were carried out using the four-state quantum model.}
	\label{fig:d40nm_two_ts_TDBC_3pc_n1_8_yz}
\end{figure}

The two resonances in \Fref{fig:TDBC_d40nm_n1_8_3p_tvary} are hybrid exciton polariton modes, denoted $\omega_-$ and $\omega_+$ for the low and high energy modes respectively. The two hybrid modes are visualized in \Fref{fig:d40nm_two_ts_TDBC_3pc_n1_8_yz} for shell thicknesses of $t=10~nm$ and $t=40~nm$. In the figure, light is presumed incident on the nanoparticle along the positive z-axis and polarised in the x-direction. The power flow is illustrated with streamlines of the time-averaged total Poynting vector $\boldsymbol{S}$ and the local value of the field enhancement is plotted on a colour scale. This figure shows that for either of the two hybridized modes there exists a region of space within which electric field enhancement occurs (bounded by the green line/area). This region of space is greater for the $\omega_+$ mode for both thicknesses considered. Conversely, the streamlines of power flow are seen to flow into the nanoparticle from a greater distance away from the nanoparticle surface for the $\omega_-$ mode in each case. Taken together, these properties of field enhancement and the nanoparticle acting as a 'magnet' for light demonstrate the validity of the consideration of these features as particle exciton-polariton modes, with similar characteristics to particle plasmon polariton modes.\\

\subsection{Dye concentration}
\begin{figure}[h]
	\centering
	\includegraphics[width=0.75\columnwidth]{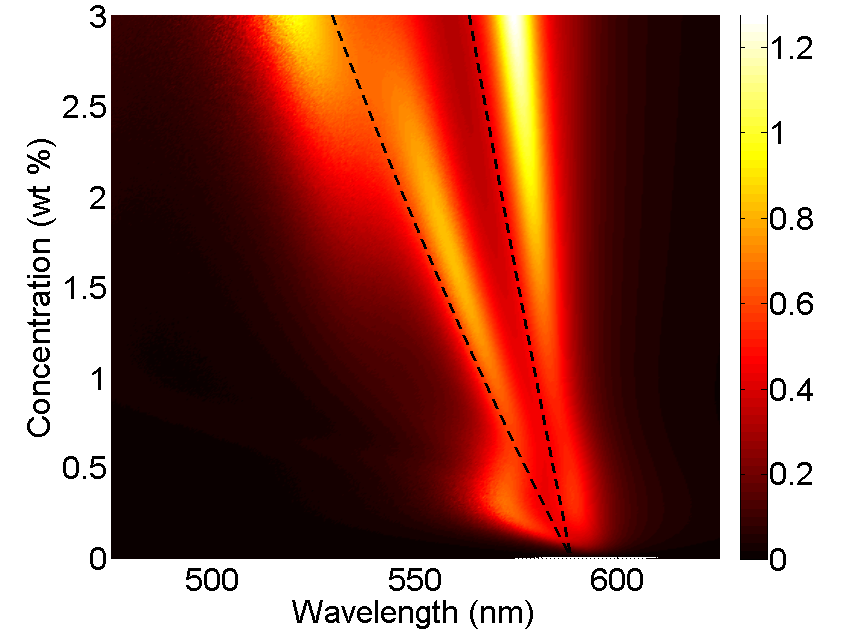}
	\caption{Colour plot: absorption efficiency spectra in vacuum as a function of concentration, for a $d=40~nm$ $t=8~nm$ nanosphere with a $n_1=2.1$ core and a TDBC:PVA shell (using the four-level quantum model for permittivity). Calculations were carried out using the four-state quantum model. Dashed lines: predicted values of the resonance frequencies using the quasistatic Lorentz oscillator model using \Eref{eq:quadratic_solns}.}
	\label{fig:TDBC_d40nm_t2nm_n1_8_concvary}
\end{figure}

The effect of the dye concentration on the splitting between the two hybrid modes is now considered. For clarity the index of the core is considered to be a constant value. Plotted in \Fref{fig:TDBC_d40nm_t2nm_n1_8_concvary} are $Q_{abs}$ spectra as a function of dye concentration in the shell for a $d=40~nm$, $t=8~nm$ nanosphere with a TDBC:PVA shell and a core with a refractive index of $n_1=2.1$. The splitting increases with concentration as predicted by the results of \Eref{eq:quadratic_solns}, visualized in \Fref{fig:TDBC_d40nm_t2nm_n1_8_concvary} by the dashed lines. The peak values in the spectra of $Q_{abs}$ also increase with concentration. For concentrations less than $1.5~wt\%$, the peak value of $Q_{abs}$ does not exceed unity for either mode. These findings suggest that, as expected, well-separated resonances require a high level of doping.\\

\subsection{Core material}
\begin{figure}[h]
	\centering
	\includegraphics[width=0.75\columnwidth]{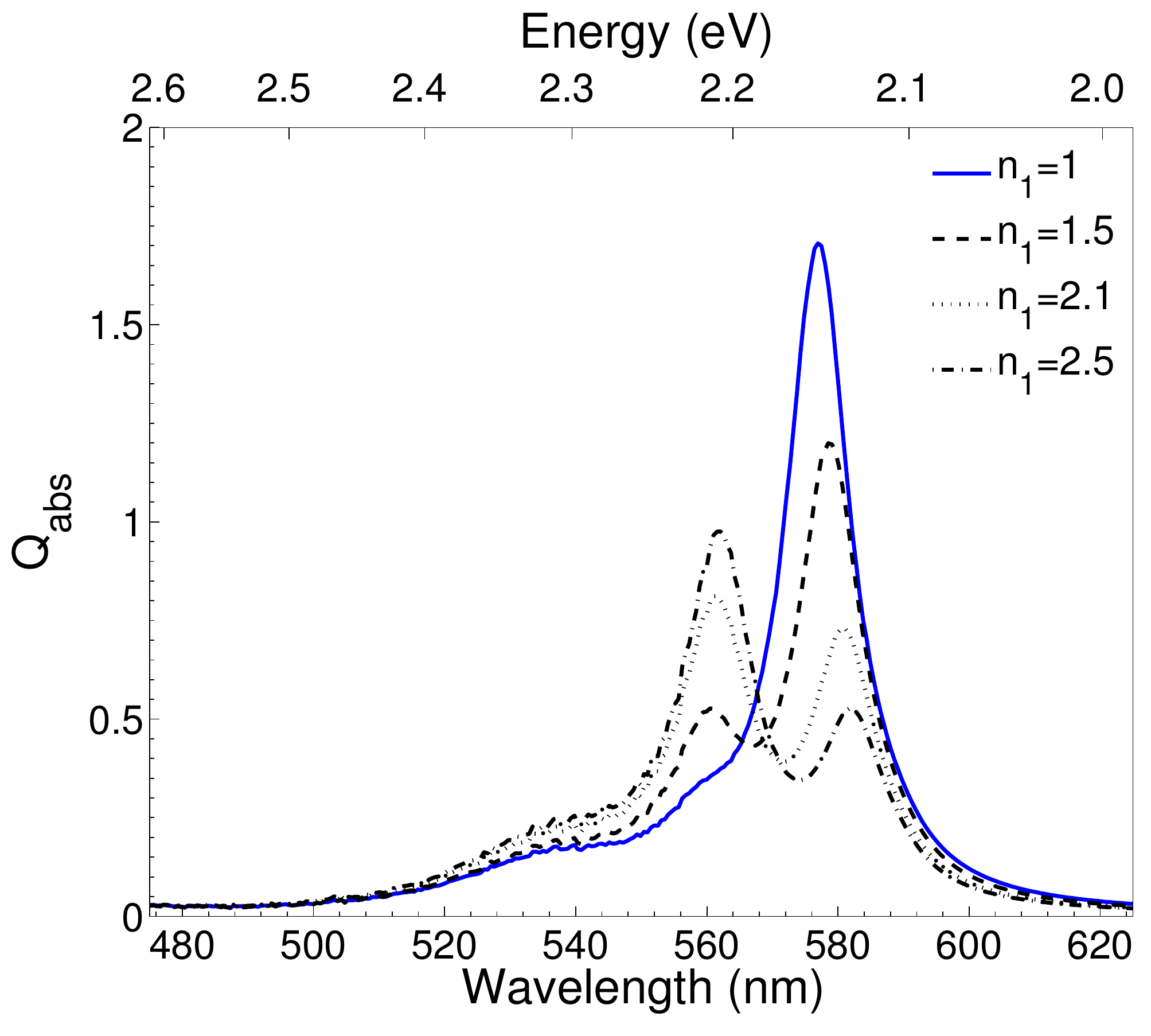}
	\caption{$Q_{abs}$ spectra for a core-shell nanosphere with a dielectric core of index $n_1$ and diameter $d=40~nm$ and a TDBC:PVA shell with dye concentration $1.5~wt\%$ and thickness $8~nm$.}
	\label{fig:Q_abs_d40nm_t10nmTDBC_3pc_on_nvary}
\end{figure}

Lastly, we discuss the effect of the core permittivity on the strength of the hybridized modes. Given that a bare non-absorbing nanosphere scatters more strongly at shorter wavelengths (Rayleigh scattering), an increase in core index will tend to increase the strength of the higher-energy mode, and \textit{vice versa}. This dependency is examined in \Fref{fig:Q_abs_d40nm_t10nmTDBC_3pc_on_nvary}, where we have plotted $Q_{abs}$ spectra for a $d=40~nm$, $t=10~nm$ nanosphere with a $1.5~wt\%$ TDBC:PVA shell for four different core indices. The choice of shell dye concentration is not altogether arbitrary, since it is chosen for a point in \Fref{fig:TDBC_d40nm_t2nm_n1_8_concvary} where the strength of the two modes are roughly in balance and free from the effects associated with the multiple energy levels at both high and low dye concentrations. For a core index of $n_1=1$, the $\omega_-$ dominates the spectrum. The two modes balance for a core index of $n_1=1.85$ (not shown), and in the range $1.5\!<n_1\!<2.1$ the two modes are both evident and well-separated. For higher core indices, the $\omega_+$ mode dominates the response, as shown for the spectrum calculated for a $n_1=2.5$ core. 

\section{Conclusions}\label{sec:conclusions}
Field enhancement has been demonstrated in principle for excitonic nanoshells. For the nanoshell geometry, there are two modes, in common with plasmonic nanoshells. As expected, we find that the frequency splitting between these two modes is maximised for thin shells and for high dye concentrations. We have shown that the extent of the splitting is very significant when compared to the linewidth of the underlying excitonic resonance. The strength of the two modes can be brought into balance by careful selection of the core index. These findings add to the increasing number of nanostructures that support interesting exciton-polariton resonances such as Tamm modes~\cite{Nunez-Sanchez_ACSPhot_2016_3_743}, and excitonic surface lattice-resonances~\cite{Humphrey_JOpt_2016_18_085004} and help to highlight the potential of these molecular nanophotonic systems. We note that hybridisation is also possible in planar (2D) molecular systems, and it might be interesting to consider whether molecular membranes, either naturally occurring or synthetic, might support excitonic-polariton modes.


\ack{The work was supported in part by the UK Engineering and Physical Sciences Research Council ((EP/K041150/1), and in part by The Leverhulme Trust.}\\

\end{document}